\documentclass[aps,prr,superscriptaddress,twocolumn,floatfix
]{revtex4-1}
\usepackage{graphicx,amsmath,mathtools,amsfonts}
\usepackage{bm}
\usepackage[normalem]{ulem}
\usepackage[usenames, dvipsnames]{xcolor}
\usepackage{multirow}
\usepackage{hyperref}
\hypersetup{colorlinks=true, linkcolor=blue, citecolor=blue, urlcolor=blue}

\usepackage[section]{placeins}

\newcommand\blue[1]{{\color{black}#1}}

\begin{document}

\title{Geometric effects in the infinite-layer nickelates}

\author{F. Bernardini}
\affiliation{Dipartimento di Fisica, Universit\`a di Cagliari, IT-09042 Monserrato, Italy}
\author{A. Bosin}
\affiliation{Dipartimento di Fisica, Universit\`a di Cagliari, IT-09042 Monserrato, Italy}
\author{A. Cano}
\affiliation{
Univ. Grenoble Alpes, CNRS, Grenoble INP, Institut Néel, 25 Rue des Martyrs, 38042, Grenoble, France
}
\date{\today}

\begin{abstract}
Geometric effects in the infinite-layer nickelates $R$NiO$_2$ associated with the relative size of the $R$-site atom are investigated via first-principles calculations. 
We consider, in particular, the prospective YNiO$_2$ material to illustrate the impact of these effects. 
\blue{Compared to LaNiO$_2$, we find that the La~$\to$~Y substitution is equivalent to a pressure of 19~GPa and that the presence of topotactic hydrogen can be precluded. However, the electronic structure of YNiO$_2$ departs from the cuprate-like picture due to an increase in both self-doping effect and $e_g$ hybridization.
}
Furthermore, we find that geometric effects introduce a quantum critical point in the $R$NiO$_2$ series. This implies a $P4/mmm \leftrightarrow I4/mcm$ structural transformation associated to a $A_3^+$ normal mode, according to which the oxygen squares undergo an in-plane rotation around Ni that alternates along $c$. 
We find that such a $A_3^+$-mode instability has a generic character in the infinite-layer nickelates and can be tuned via either the effective $R$-site atom size or epitaxial strain. 

\end{abstract}

\maketitle

\section{Introduction}

The infinite-layer nickelates $R$NiO$_2$ ($R=$ rare-earth element) have long been discussed as candidate materials for cuprate-like high-$T_c$ superconductivity \cite{hayward99,anisimov99,pickett-prb04,norman20-p}. Thus, the recent discovery of unconventional superconductivity in these systems has generated a substantial research attention \cite{cano-review,tao-review,arita21review}.  
Initially, superconductivity was observed in Sr-doped thin films for $R=$~Nd but not for $R=$~La \cite{hwang19a}. This circumstance has raised the question of the role of the 
4$f$ electrons \cite{choi20_1} as well as of extrinsic factors such as the unintentional presence of topotatic hydrogen or apical oxygens, which have been argued to correlate with the structure \cite{held20topotaticH,cano20c}. 
Examining the $R$NiO$_2$ series for $R$ from La to Lu as well as additional candidate materials for nickelate superconductivity has thus emerged as a consequential outlook \cite{hwang21-La,ariando21-La,hwang20Pr-a,schilling22-njphy,sm-112,botanaprm,botana20magnetism,jia-prx21,arita20prb,cano20d,cano21-prm,higher_order,mundy21}. 
At the same time, in analogy with their perovskite counterparts \cite{ghosez17,bibes17,medarde19}, the appearance of structural instabilities associated to the relative size of the $R$ atom may also come into play 
as an additional key ingredient. To the best of our knowledge, this latter possibility remains unexplored so far. 

In this context, YNiO$_2$ provides a prospective model-case system combining the absence of 4$f$ electrons and geometric effects associated with a comparatively small $R$ atom. Here, we introduce this system as a proxy to analyze the impact of these effects on the fundamental properties of the infinite-layer nickelates. 
Specifically, by means of first-principle calculations, 
we quantify the corresponding changes in both crystal structure and electronic properties, addressing also the likeliness of topotactic hydrogen insertion. Our analysis of geometrical effects further reveals the presence of a structural quantum critical point across the infinite-layer rare-earth nickelate series. We show that this structural instability is related to a distinct A$_3^+$ soft-mode behavior that can be tuned via either the effective size of the atom at the $R$ site or epitaxial strain.  

\section{Computational methods}

The equilibrium structure and phonon calculations were performed using the {\sc Quantum ESPRESSO} code \cite{QE}. \blue{
We used norm-conserving PBE and PBEsol pseudopotentials from the Dojo library with a 115 (460) Ry cutoff for wavefunction (density) \cite{ONCVPSP,dojo}. 
In addition, we also used the PBEsol pseudo-potentials from the SSSP library with a 75 (600) Ry cutoff for wavefuntions (density) ~\cite{PBEsol,SSSP}. 
} 
The lattice parameters obtained for LaNiO$_2$ and YNiO$_2$ with these two sets of PBEsol pseudopotentials are essentially identical. However, significant differences are obtained with PBE vs PBEsol as discussed \blue{below}. 
In the case of the Pr - Eu elements, we restrict ourselves to \blue{the Dojo} pseudopotentials in which the 4$f$ electrons are treated as core electrons. 
The phonon dispersion was computed using DFPT~\cite{DFPT}. The Brillouin zone integration was performed using a $16\times16\times20$ $k$-mesh and a $8\times 8\times 10$ $q$-mesh was used to compute the dynamical matrix. Further, the optimized structural parameters were employed to compute the electronic structure properties using the all-electron code {\sc{WIEN2k}}~\cite{WIEN2k} based on the full-potential augmented plane-wave plus local orbitals method (APW+LO). 
We used muffin-tin radii of 2.5, 2.10, and 1.48 a.u. for the Y, Ni, and O atoms, respectively, and a plane-wave cutoff $R_{\rm MT}K_{\rm max}$ = 7.0. 
The integration over the Brillouin zone was done using a Monkhorst-Pack mesh of $11 \times 11 \times 14$ $k$-points for the self-consistent calculations.

In addition, we performed magnetic calculations using both PBEsol and the local density approximation (LDA)~\cite{LDA} as the 
\blue{generalized gradient approximation (GGA) tends} 
to overestimate the tendency towards magnetic order in 
\blue{weakly and moderately correlated} metals \cite{mazin08-prb}. The magnetic solutions were obtained using different cells according to the magnetic order under consideration. 
For the ferromagnetic (FM) order we used the primitive tetragonal cell, while for the $A$-type
antiferromagnetic (AFM) order with the spins of adjacent NiO$_2$ planes pointing in opposite directions ---{\it i.e.} FM planes stacked AFM along $c$--- we used a tetragonal cell encompassing 2 formula units. 
For the $C$-type AFM order ---{\it i.e.} checkerboard order in-plane --- 
we used a $\sqrt{2}\times\sqrt{2}\times{1}$ tetragonal cell encompassing 2 formula units, with $a$ and $b$-axis rotated by 45$^\circ$ with respect to the primitve cell.
For the $G$-type AFM with nearest–neighboring spins pointing in opposite directions (``full" AFM ordering) we used a $\sqrt{2}\times\sqrt{2}\times2$ body-centered tetragonal cell.
Finally, we also considered the $E$-type AFM order where spins are aligned forming a double stripe structure using a base-centered orthorhombic cell encompassing 4 formula units.

\section{Results}

\subsection{Structure}

We first consider the ideal $P4/mmm$ structure of the prospective nickelate YNiO$_2$. 
\blue{Using the PBEsol functional,} the optimized lattice parameters are 
$a=3.798$~{\AA} and $c=3.092$~{\AA}.
Compared to LaNiO$_2$, the volume of the unit cell is reduced due to the smaller size of the Y cation. 
In fact, the $a$ parameter is smaller compared to the values reported across the $R$NiO$_2$ series ($R =$~La - Lu) while the parameter $c$ is similar to the value calculated for ErNiO$_2$ \cite{jia-prx21,botana20magnetism}. 
We note that this difference, however, is resolved when the PBE functional is used instead. In that case, we obtain $a=3.850$~{\AA} and $c=3.144$~{\AA}, which places YNiO$_2$ right between TbNiO$_2$ and ErNiO$_2$ in terms of both $a$ and $c$ lattice parameters. 
\blue{We note that, for the experimentally available compounds ($R =$~La, Pr, and Nd), PBEsol performs remarkably well in reproducing the $c$ parameter but not $a$ and, conversely, PBE performs very well with $a$ but not $c$ (see Table \ref{t:structure} and  \cite{hepting21-sciadv,mitchell20bulk,schilling22-njphy}). 
The overall agreement (lattice constants and anisotropy), however, is slightly better with PBE. 
}

\blue{
In order to translate this reduction of lattice parameters into an effective pressure we computed the bulk modulus of LaNiO$_2$ using PBE. The calculated value is 150~GPa, which is similar to that measured in CaCuO$_2$ and other cuprates (see e.g. \cite{jin05-bulkmodulus}). Thus, from the lattice parameters obtained for YNiO$_2$, the La~ $\to$~Y substitution in these infinite-layer nickelates can be seen as equivalent to a chemical pressure of $\sim$~19 GPa. Taking into account that LaNiO$_2$ has been reported to be superconducting with $T_c \sim $~1\,K and that the $T_c$ of (Pr,Sr)NiO$_2$ has been reported to increase with pressure ---without saturation up to 12~GPa--- at a rate $dT_c/dP \sim 1$~K/GPa \cite{cheng21-pressure}, it is tempting to speculate that YNiO$_2$ may be superconducting with $T_c\sim$19~GPa. 
}

\subsection{Topotactic hydrogen}

Infinite-layer nickelates are typically obtained from their perovskite counterparts by means of topochemistry, using reducing agents such as CaH$_2$. This process, however, may lead to oxide-hydrides $R$NiO$_2$H instead, as concluded for $R=$~La but not for Nd after partial substitution with Sr \cite{held20topotaticH}.
This suggests a link between the presence of topotactic hydrogen and the size of the rare-earth element that can be further verified by analyzing YNiO$_2$.

\begin{table}[b!]
\begin{tabular}{c p{1ex} c c c c}
\hline \hline 
H position && $a$ (\AA) & $b$ (\AA) & $c$ (\AA) & $\Delta E$ (eV/f.u.)\\
\hline
Apical 
&& 3.850   & 3.850 & 3.180 &     0\\
Interstitial 
&& 4.076   &3.734 & 3.236 & +2.38\\
Equatorial 
&& 3.830   & 3.830 & 3.486 & +3.25\\
\hline \hline
\end{tabular}
\caption{
Lattice parameters and relative energy of YNiO$_2$H calculated for different positions of the H atom \blue{in the $P4/mmm$ structure using the PBEsol functional.
}
}
\label{t:topoH}
\end{table}

\begin{table}[b!]
\blue{
\begin{tabular}{c p{2ex} c p{2ex} c c }
\hline \hline 
\multirow{2}{*}{$U$ (eV)}&& \multicolumn{3}{c}{$E_B$ (meV/f.u.)} \\
&& $P4/mmm$  && $I4/mcm$ \\
\hline
0 &&   +316 (+197) && +57 \;($-$38) \\
1 &&   +339 (+217) && +79 \;($-$20) \\
2 &&   +366 (+242) && +104 (+2 ) \\
3 &&   +397 (+269) && +133 (+27) \\
4 &&   +432 (+300) && +165 (+55) \\ 
5 &&   +471 (+335) && +201 (+86) \\
\hline \hline
\end{tabular}
}
\caption{
\blue{
Binding energy $E_B$ of the topotatic H at the apical position obtained from PBE+$U$ calculations as a function of the Hubbard $U$ parameter. The values in parenthesis correspond to the PBEsol functional. 
}
}
\label{t:topoH+U}
\end{table} 

Thus, to determine whether the nickelates with smaller $R$ atoms may host topotactic hydrogens, we first consider YNiO$_2$H with three possible configurations \blue{relative to the $P4/mmm$ strucure}. Namely, with the H atom occupying either apical, interstitial or equatorial positions. The computed structures and relative energies are given in Table \ref{t:topoH}. 
We find that the configuration in which the H sits at the apical positions is energetically favoured. We note that the same situation takes place in LaNiO$_2$ \cite{held20topotaticH}.
The H binding energy $E_B$ can then be computed from the corresponding total energies as \cite{held20topotaticH} 
   \begin{align}
       \blue{ 
        E_B =  E[\text{YNiO$_2$H}] - 
        \big( E[\text{YNiO$_2$}] + E[\text{H$_2$}]/2\big).
        }
    \end{align}
Thus, we find that, contrary to LaNiO$_2$, the incorporation of H into YNiO$_2$ is energetically unfavorable by \blue{197~meV/f.u. using the PBEsol functional and even more unfavorable by 316~meV/f.u. using the PBE one. 
Besides, we find that correlation effects at the DFT+$U$ level penalize even further the H incorporation (see Table \ref{t:topoH+U}).} 
This is in fact in tune with the trend previously obtained as a function of Sr-doping and strain \cite{held20topotaticH}. 
Our results thus confirm that the $R$NiO$_2$ nickelates are borderline systems where the H binding energy can change from positive to negative. In the case of (La$\to$Y)NiO$_2$, topotactic hydrogen is avoided by means of a Y$^{3+}$ cation that preserves the nominal 1+ oxidation state of the Ni atom.

\subsection{Electronic properties}

\begin{figure}[t!]
\includegraphics[width=.45\textwidth]{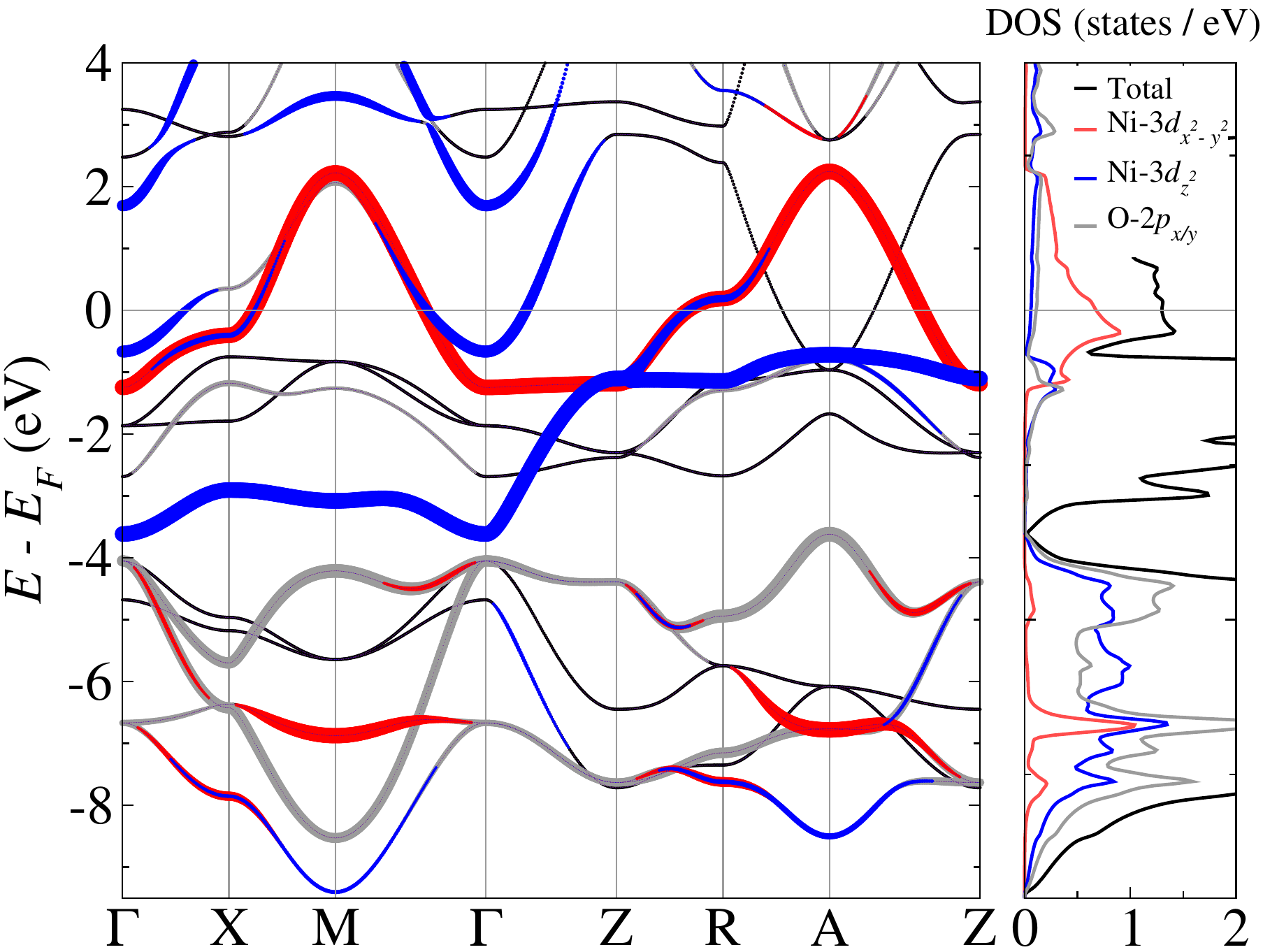}
    \caption{Band structure along the high-symmetry directions of the Brillouin zone and density of states of YNiO$_2$ in the $P4/mmm$ structure.}
    \label{f:bands-dos}
\end{figure}

Fig. \ref{f:bands-dos} shows the band structure and orbital-resolved density of states (DOS) of YNiO$_2$ in the $P4/mmm$ structure.
The main features near the Fermi level are associated to the Ni-3$d$ bands. These features can be traced back to the nominal 3$d^9$ configuration of the Ni$^{1+}$ as in LaNiO$_2$. 
The splitting between the $e_g=\{d_{x^2-y^2},d_{z^2}\}$ states, however, is considerably reduced. 
At the same time, there is a sizable self-doping from the ideal $d^9$ configuration which, despite the reduced polarizability of the Y ion, is visibly enhanced compared to LaNiO$_2$. 
Such a self-doping originates from a Y-4$d$ derived band, which has an increased hybridization not only with Ni-3$d_{z^2}$ but also with O-2$p$ states. In fact the width of the O-2$p$ bands increases which makes the contribution of the O-2$p$ states to the DOS near the Fermi level slightly higher compared to LaNiO$_2$. These modifications result from the reduced lattice parameters of YNiO$_2$, and are compatible with those reported in \cite{botana20magnetism} for LaNiO$_2$ as a function of the in-plane lattice constant. Also, the presence of the two larger electron pockets at $\Gamma$ and A is in tune with those reported in \cite{jia-prx21}.

These changes in the electronic structure modify the tendency towards magnetism as summarized in Table \ref{t:magnetism}. Specifically, our calculations with FM and $A$-AFM initial configurations converged but yielded a vanishing magnetic moment \blue{while the} rest of configurations resulted in metastable states at the LDA level. 
\blue{This situation changes at the GGA level. In this case, the tendency towards magnetism is enhanced and the $G$-AFM solution is found to minimize the energy. 
In any case, we note that both LDA and GGA yield rather small magnetization energies indicating that the tendency towards magnetism is very weak.}

\subsection{$P4/mmm \to I4/mcm$ structural distortion \& epitaxial strain} 

Next, we address the stability of the $P4/mmm$ structure of the infinite-layer $R$NiO$_2$ nickelates when the size of the $R$ atom is reduced. 
The Goldschmidt tolerance factor is a well known geometric indicator of the stability of cubic perovskites. 
In the tetragonal infinite-layer case, there is an analogous `no rattle limit' that can be inferred from the ratio of ionic radii also. Accordingly, for $R=$~Y, we note that the ionic-radius ratio $r_{\rm Y}/r_{\rm O}$ in 8-fold coordination is $0.718$, and hence below such a limit (0.732).
In fact, the calculated Y-O distance in YNiO$_2$ is slightly larger than the sum of the corresponding radii.
Consequently, the system may be prone to structural instabilities according to Pauling's rules \cite{pauling60}. 

\begin{table}[t!]
\begin{tabular}{r c c c c}
\hline \hline 
 && $\Delta E$ (meV/Ni) && $\mu_{\rm Ni}$ ($\mu_B$) \\
\hline
$G$-AFM && $+$2.98 \blue{($-$9.16)} && 0.30 \blue{(0.45)} \\
$E$-AFM && $+$1.18 \blue{($-$3.65)} && 0.14 \blue{(0.37)}\\
$C$-AFM && $+$0.27 \blue{($-$1.92)} && 0.06 \blue{(0.43)}\\
$A$-AFM && - && - \\
FM && - && - \\
\hline \hline
\end{tabular}
\caption{
Energy difference with respect to the non-magnetic solution and magnetic moment of the Ni atom for different magnetic configurations of YNiO$_2$ obtained \blue{using the LDA exchange-correlation functional (the values between parenthesis correspond to PBEsol)}. 
}
\label{t:magnetism}
\end{table}

\begin{figure}[b!]
\includegraphics[width=.375\textwidth]{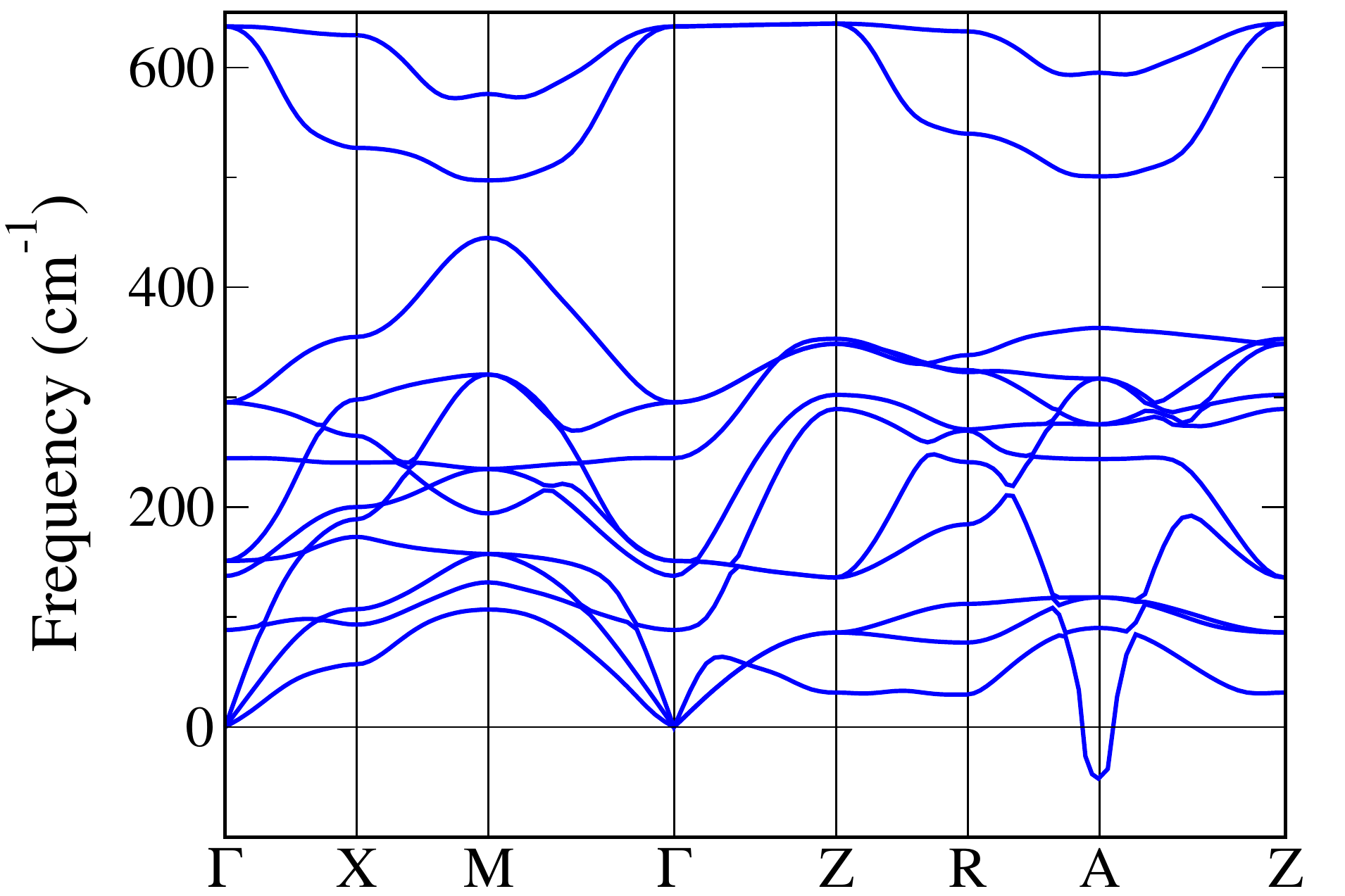}
    \caption{Phonon spectrum of YNiO$_2$ computed in the optimized $P4/mmm$ structure. The spectrum reveals a weak instability at A [$\mathbf q=(1/2,1/2,1/2$)] associated to an $A_3^+$ normal mode. }
    \label{f:phonons}
\end{figure}

\begin{figure}[t!]
\includegraphics[width=.475\textwidth]{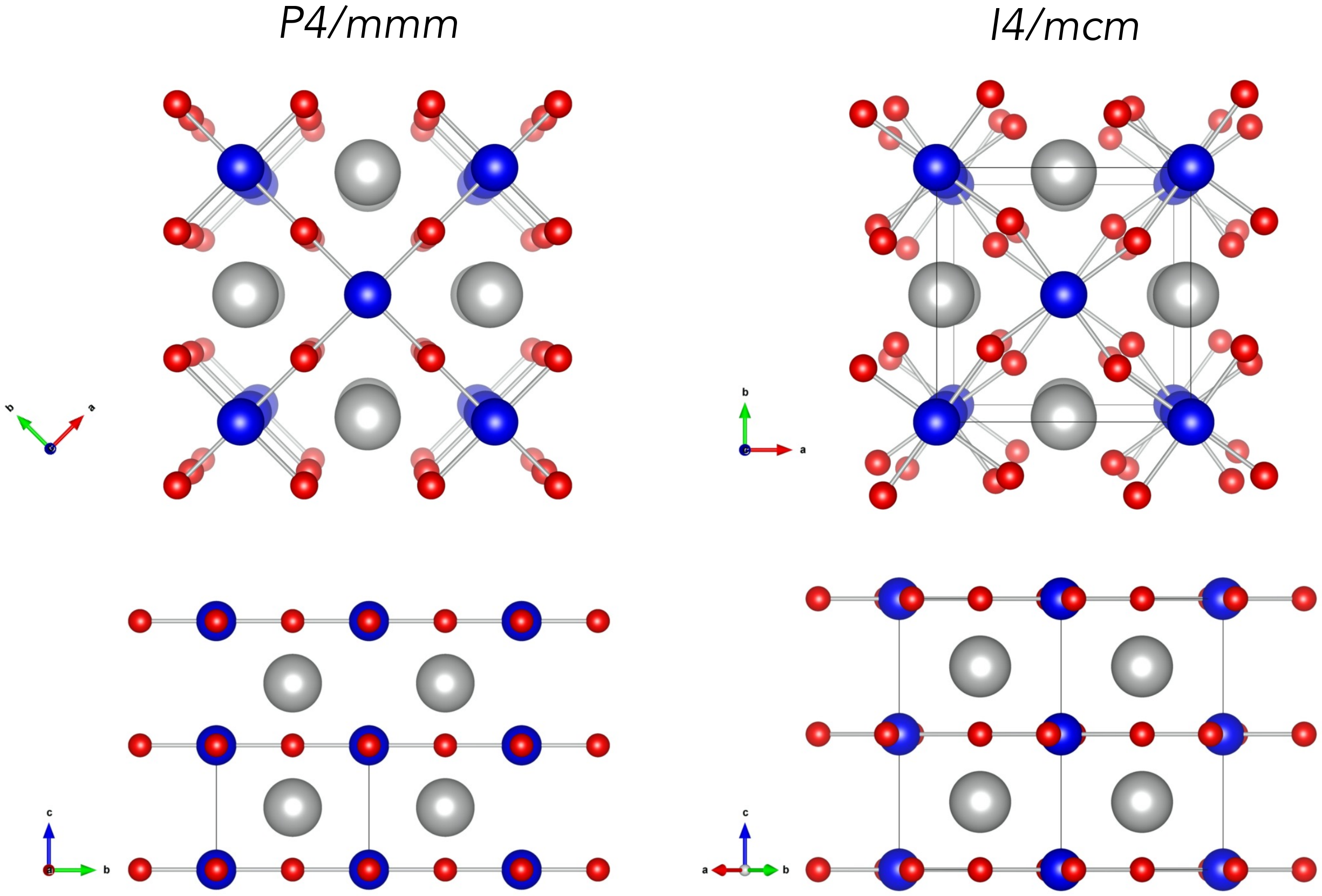}
    \caption{
    Top and side views of the ideal $P4/mmm$ and distorted $I4/mcm$ structures of the infinite-layer nickelates $R$NiO$_2$. Gray, blue and red balls indicate $R$, Ni and O atoms respectively.}
    \label{f:structures}
\end{figure}

To confirm this, we considered possible distortions of the $P4/mmm$ structure. 
The phonon spectrum, in particular, reveals a dynamical instability at the A point of the Brilluoin zone (see Fig. \ref{f:phonons}). 
In fact, we further find that the total energy is minimized in the $I4/mcm$ structure illustrated in Fig. \ref{f:structures}. The optimized lattice parameters \blue{again depend on the exchange-correlation functional that is used. Using PBEsol we find} $a=5.390 $~{\AA} and $c=6.376$~{\AA}, with the parameter $x_{\rm O} = 0.03873$ determining the oxygen positions at the 8$h$ Wyckoff sites.
\blue{Using PBE, however, we find $a=5.480 $~{\AA}, $c=6.499$~{\AA}, and $x_{\rm O} = 0.04528$.
}

\blue{
When it comes to the possibility of incorporating topotactic H, the $I4/mcm$ structure becomes formally borderline. In this case, the calculated binding energy ranges between $-$38 meV/f.u. using PBEsol and $+$57 meV/f.u. using PBE. However, taking into account electronic correlations at the DFT+$U$ level the binding energy further increases and eventually becomes positive in both cases (see Table \ref{t:topoH+U}). This indicates that the presence of H is also precluded in this structure. 
Moreover, note that these values correspond to zero temperature while, experimentally, the topotactic reduction is typically performed at $\sim$~$300^\circ$C. At this temperature, the $I4/mcm$ distortion can be expected to be significantly weaker so that the binding energy will tend to that obtained for the $P4/mmm$ case. 
}

As discussed above, this $P4/mmm \to I4/mcm$ instability can be ascribed to the presence of an $R$ atom that is too small (i.e. a geometric effect). In fact, we find a gradual softening of the A$_3^+$ phonon across the series that eventually reveals the dynamical instability of the $P4/mmm$ structure as summarized in Table \ref{t:structure}. Specifically, while the $R=$~La, Pr, and Nd members remain stable, the structural instability emerges for smaller $R$ elements. At the same time, we obtain different A$_3^+$-mode frequencies depending on the exchange-correlation ---PBE vs PBEsol--- functional that is employed. Thus, while the existence of a critical $R$ size is clear from the calculated trend, we could not conclude its precise value.

\begin{figure}[b!]
    \includegraphics[width=.35\textwidth]{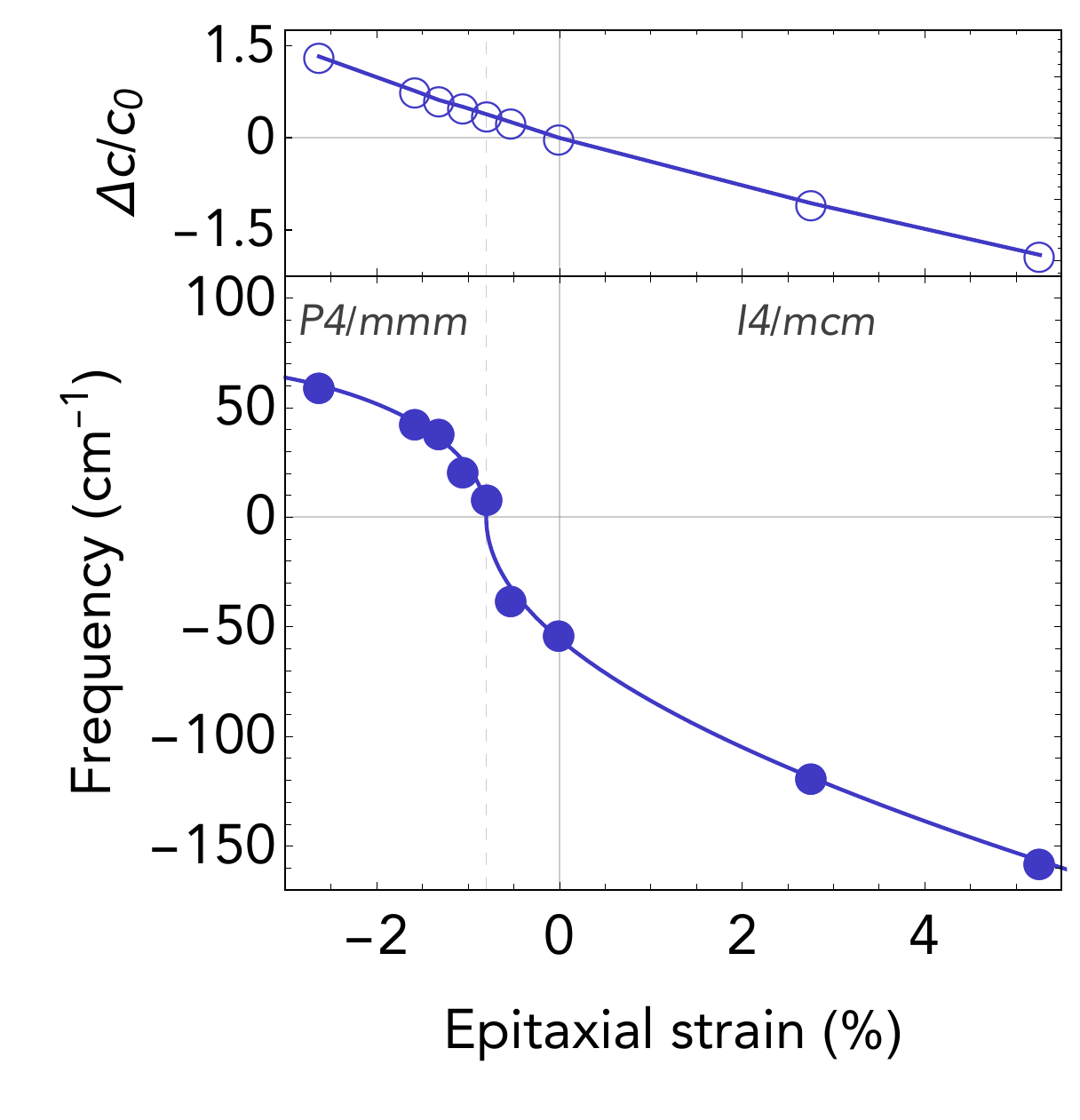}
    \caption{\blue{Relative change in the} out-of-plane lattice parameter $c$ (top panel) and frequency of the soft $A_3^+$ mode (bottom panel) of YNiO$_2$ calculated in the $P4/mmm$  structure as a function of epitaxial strain. Negative values correspond to imaginary frequencies indicating the dynamical instability towards the $I4/mcm$ structure. Compressive strain $\lesssim -0.7$~\% suppresses \blue{the} $A_3^+$-mode instability.
    }
    \label{fig:structure-vs-strain}
\end{figure}

In any case, the structural instability of the $P4/mmm$ phase is relatively weak. In fact, for YNiO$_2$, the difference in energy between the $P4/mmm$ and the $I4/mcm$ structures is only 45~meV per formula unit. Thus, we considered 
the influence of epitaxial strain since this may prevent the in-plane oxygen rotations within the NiO$_2$ layers. 
Epitaxial strain $\epsilon$ is modeled by changing the in-plane lattice parameter and relaxing the structure in the $c$ direction [$\epsilon=(a-a_0)/a_0$, where $a_0$ is the initial (relaxed) lattice constant]. 
The top panel in Fig. \ref{fig:structure-vs-strain} shows the calculated $c$ parameter as a funcion of epitaxial strain in the $P4/mmm$ structure. The $c$ parameter increases by decreasing the lattice constant $a$, and in this sense counteracts the La $\to$ Y substitution. 
This behavior is standard and can be expected across the $R$NiO$_2$ series also. 
The overall modification of the structure, however, remains dominated by the relative changes imposed in $a$. Consequently, compressive strain can be expected to reduce the Y `rattle' and thereby to stabilize the $P4/mmm$ structure.

\begin{table}[t!]
\begin{tabular}{c c c c c c c}
\hline \hline 
   && $a$ (\AA) && $c$ (\AA) && $\omega_{\rm A_3^+}$~(cm$^{-1}$) \\
\hline 
La && 3.939 (3.880) && 3.398 (3.352) && 132.0 (178.0)\\
Pr && 3.920 \blue{(3.864)} && 3.355 \blue{(3.314)} && 119.7 \blue{(169.6)} \\
Nd && 3.905 \blue{(3.849)}&& 3.310 \blue{(3.267)} &&  95.2 \blue{(153.9)} \\
Sm && 3.880 \blue{(3.826)} && 3.237 \blue{(3.191)} && $i$38.1  \blue{(115.0)}\\
Eu && 3.870 \blue{(3.816)}&& 3.207 \blue{(3.160)}&& $i$82.6 \blue{(89.6)} \\
Y  && 3.850 (3.798) && 3.144 (3.092) && $i$130.1 ($i$52.5) \\
\hline \hline
\end{tabular}
\caption{
Optimized lattice parameters and frequency of the A$_3^{+}$ mode for different $R$NiO$_2$ systems in the $P4/mmm$ structure obtained with the PBE exchange-correlation functional (the values between brackets correspond to the PBEsol functional). 
Imaginary frequencies indicate the dynamical instability of the $P4/mmm$ structure.}
\label{t:structure}
\end{table}

The bottom panel of Fig. \ref{fig:structure-vs-strain} shows the computed frequency of the soft A$_3^+$ mode as a function of epitaxial strain. 
As we see, this mode hardens for compressive strain so that the corresponding instability disappears for $\epsilon \lesssim -0.7$~\%. 
This confirms that the in-plane compression obtained via epitaxial strain overcomes the out-of-plane expansion, thereby providing a direct handle to stabilize the $P4/mmm$ structure. 
\blue{
This could be achieved by using substrates such as $R$AlO$_3$ ($R =$~La, Nd, Y and Lu).}
Conversely, we find that tensile strain enhances the A$_3^+$-mode instability. We have performed additional calculations for other $R$NiO$_2$ systems. These calculations reveal that this soft-mode behavior as a function of epitaxial strain is in fact a generic feature of the infinite-layer rare-earth nickelates. Accordingly, the $I4/mcm$ structure with the O-square rotations should be expected even in LaNiO$_2$ under high enough tensile strain.

\FloatBarrier

\section{Summary}

In summary, we have investigated geometric effects in the infinite-layer rare-earth nickelates by means of first principles calculations. 
By considering the prospective material YNiO$_2$, we have confirmed that 
\blue{cation-size} 
effects can have a decisive influence on the actual reduction of their perovskite precursors since they can prevent the presence of topotactic hydrogen. 
\blue{Moreover, the La~$\to$~Y substitution is equivalent to a pressure of 19~GPa, which further might be favorable for superconductivity.} 
However, the price to pay is the substantial modification of the electronic properties, which qualitatively deviate from the ideal cuprate-like case. Specifically, both the self-doping and the $e_g$ hybridization ---{\it i.e.}, the 3D character of the system--- is enhanced while tendency towards magnetic order is weakened. Furthermore, we have shown that there exists a structural quantum critical point underneath the $R$NiO$_2$ series. Specifically, we have found that the original $P4/mmm $ structure eventually becomes unstable by either reducing the size of the $R$ atom or under high-enough epitaxial tensile strain. This gives rise to a new $I4/mcm$ superstructure with spontaneous, in-plane rotations of the oxygen squares according to an A$_3^+$ soft-mode distortion. 
Interestingly, this structural instability is fundamentally different from its counterpart in the perovskite nickelates where the A$_3^+$ mode is totally passive without contributing to their overall distortion \cite{bibes17,ghosez17,medarde19}. 
In the infinite-layer nickelates, where the apical oxygens are absent, the A$_3^+$-mode becomes the main driver of their 
$P4/mmm \leftrightarrow I4/mcm$ 
structural transformation instead. At the same time, we note that this instability is also different from the out-of-plane oxygen buckling reported for CaCuO$_2$ \cite{liechtenstein96}. 
Thus, our work establishes new links between geometric effects and fundamental properties of the infinite-layer nickelates and we expect that our findings will motivate further investigations. 

\blue{
\emph{Note added.} 
After the submission of this work, there appeared three preprints addressing similar questions from first-principles calculations. The soft-mode instability and its dependence on epitaxial strain across the $R$NiO$_2$ series ($R = $~La - Lu) is confirmed in \cite{chen21-dyn,ghosez21}. 
An additional $Pbmn$ structure is reported in \cite{bibes21}. 
The relative stability of this structure in YNiO$_2$, however, depends on the exchange-correlation functional that is used. 
With respect to the $I4/mcm$ phase, $\Delta E = -27$, $-$57 and $-37$~meV/f.u. is reported using PBEsol, PBEsol+$U$ and SCAN respectively in \cite{bibes21} while we find $\Delta E = -41$, $+6$ and $-25$~meV/f.u. using PBE, PBEsol and PBEsol+$U$ ($U = 5$~eV on the Ni in both cases). Also, the binding energy of topotactic H in the $I4/mcm$ structure is reported to be $-$300~meV/f.u. using SCAN in \cite{bibes21} while we find $-$38 and $+$57 meV/f.u. using PBEsol and PBE respectively. We note that, when it comes to the structural parameters of the infinite-layer nickelates, PBE and PBEsol perform considerably better than SCAN in non-magnetic calculations \cite{zhang21-commphys}. On the other hand, when it comes to magnetism, the computed magnetic transition temperatures are always below 160~K \cite{ghosez21}, so that the system can be expected to be paramagnetic at the reduction temperature where the hypothetical H incorporation may take place ({\it i.e.} at $\gtrsim 500$~K). 
}

\section*{Acknowledgments} 
F.B. acknowledges support from Cineca ISCRA-C project "IsC78-NICKSUP-HP10C91RDL". A.C. acknowledeges support from the Visiting Professor/Scientist 2019 program founded by the Regione Autonoma Sardegna.  
\bibliography{bib.bib}

\begin{thebibliography}{46}%
\makeatletter
\providecommand \@ifxundefined [1]{%
 \@ifx{#1\undefined}
}%
\providecommand \@ifnum [1]{%
 \ifnum #1\expandafter \@firstoftwo
 \else \expandafter \@secondoftwo
 \fi
}%
\providecommand \@ifx [1]{%
 \ifx #1\expandafter \@firstoftwo
 \else \expandafter \@secondoftwo
 \fi
}%
\providecommand \natexlab [1]{#1}%
\providecommand \enquote  [1]{``#1''}%
\providecommand \bibnamefont  [1]{#1}%
\providecommand \bibfnamefont [1]{#1}%
\providecommand \citenamefont [1]{#1}%
\providecommand \href@noop [0]{\@secondoftwo}%
\providecommand \href [0]{\begingroup \@sanitize@url \@href}%
\providecommand \@href[1]{\@@startlink{#1}\@@href}%
\providecommand \@@href[1]{\endgroup#1\@@endlink}%
\providecommand \@sanitize@url [0]{\catcode `\\12\catcode `\$12\catcode
  `\&12\catcode `\#12\catcode `\^12\catcode `\_12\catcode `\%12\relax}%
\providecommand \@@startlink[1]{}%
\providecommand \@@endlink[0]{}%
\providecommand \url  [0]{\begingroup\@sanitize@url \@url }%
\providecommand \@url [1]{\endgroup\@href {#1}{\urlprefix }}%
\providecommand \urlprefix  [0]{URL }%
\providecommand \Eprint [0]{\href }%
\providecommand \doibase [0]{http://dx.doi.org/}%
\providecommand \selectlanguage [0]{\@gobble}%
\providecommand \bibinfo  [0]{\@secondoftwo}%
\providecommand \bibfield  [0]{\@secondoftwo}%
\providecommand \translation [1]{[#1]}%
\providecommand \BibitemOpen [0]{}%
\providecommand \bibitemStop [0]{}%
\providecommand \bibitemNoStop [0]{.\EOS\space}%
\providecommand \EOS [0]{\spacefactor3000\relax}%
\providecommand \BibitemShut  [1]{\csname bibitem#1\endcsname}%
\let\auto@bib@innerbib\@empty
\bibitem [{\citenamefont {Hayward}\ \emph {et~al.}(1999)\citenamefont
  {Hayward}, \citenamefont {Green}, \citenamefont {Rosseinsky},\ and\
  \citenamefont {Sloan}}]{hayward99}%
  \BibitemOpen
  \bibfield  {author} {\bibinfo {author} {\bibfnamefont {M.}~\bibnamefont
  {Hayward}}, \bibinfo {author} {\bibfnamefont {M.}~\bibnamefont {Green}},
  \bibinfo {author} {\bibfnamefont {M.}~\bibnamefont {Rosseinsky}}, \ and\
  \bibinfo {author} {\bibfnamefont {J.}~\bibnamefont {Sloan}},\ }\href@noop {}
  {\bibfield  {journal} {\bibinfo  {journal} {Journal of the American Chemical
  Society}\ }\textbf {\bibinfo {volume} {121}},\ \bibinfo {pages} {8843}
  (\bibinfo {year} {1999})}\BibitemShut {NoStop}%
\bibitem [{\citenamefont {Anisimov}\ \emph {et~al.}(1999)\citenamefont
  {Anisimov}, \citenamefont {Bukhvalov},\ and\ \citenamefont
  {Rice}}]{anisimov99}%
  \BibitemOpen
  \bibfield  {author} {\bibinfo {author} {\bibfnamefont {V.~I.}\ \bibnamefont
  {Anisimov}}, \bibinfo {author} {\bibfnamefont {D.}~\bibnamefont {Bukhvalov}},
  \ and\ \bibinfo {author} {\bibfnamefont {T.~M.}\ \bibnamefont {Rice}},\
  }\href {\doibase 10.1103/PhysRevB.59.7901} {\bibfield  {journal} {\bibinfo
  {journal} {Phys. Rev. B}\ }\textbf {\bibinfo {volume} {59}},\ \bibinfo
  {pages} {7901} (\bibinfo {year} {1999})}\BibitemShut {NoStop}%
\bibitem [{\citenamefont {Lee}\ and\ \citenamefont
  {Pickett}(2004)}]{pickett-prb04}%
  \BibitemOpen
  \bibfield  {author} {\bibinfo {author} {\bibfnamefont {K.-W.}\ \bibnamefont
  {Lee}}\ and\ \bibinfo {author} {\bibfnamefont {W.~E.}\ \bibnamefont
  {Pickett}},\ }\href {\doibase 10.1103/PhysRevB.70.165109} {\bibfield
  {journal} {\bibinfo  {journal} {Phys. Rev. B}\ }\textbf {\bibinfo {volume}
  {70}},\ \bibinfo {pages} {165109} (\bibinfo {year} {2004})}\BibitemShut
  {NoStop}%
\bibitem [{\citenamefont {Norman}(2020)}]{norman20-p}%
  \BibitemOpen
  \bibfield  {author} {\bibinfo {author} {\bibfnamefont {M.~R.}\ \bibnamefont
  {Norman}},\ }\href@noop {} {\bibfield  {journal} {\bibinfo  {journal}
  {Physics}\ }\textbf {\bibinfo {volume} {13}},\ \bibinfo {pages} {85}
  (\bibinfo {year} {2020})}\BibitemShut {NoStop}%
\bibitem [{\citenamefont {Botana}\ \emph {et~al.}(2021)\citenamefont {Botana},
  \citenamefont {Bernardini},\ and\ \citenamefont {Cano}}]{cano-review}%
  \BibitemOpen
  \bibfield  {author} {\bibinfo {author} {\bibfnamefont {A.~S.}\ \bibnamefont
  {Botana}}, \bibinfo {author} {\bibfnamefont {F.}~\bibnamefont {Bernardini}},
  \ and\ \bibinfo {author} {\bibfnamefont {A.}~\bibnamefont {Cano}},\ }\href
  {\doibase 10.1134/S1063776121040026} {\bibfield  {journal} {\bibinfo
  {journal} {JETP}\ }\textbf {\bibinfo {volume} {159}},\ \bibinfo {pages} {711}
  (\bibinfo {year} {2021})},\ \Eprint {http://arxiv.org/abs/2012.02764}
  {arXiv:2012.02764} \BibitemShut {NoStop}%
\bibitem [{\citenamefont {{Zhang}}\ and\ \citenamefont {{Tao}}()}]{tao-review}%
  \BibitemOpen
  \bibfield  {author} {\bibinfo {author} {\bibfnamefont {J.}~\bibnamefont
  {{Zhang}}}\ and\ \bibinfo {author} {\bibfnamefont {X.}~\bibnamefont
  {{Tao}}},\ }\href@noop {} {\ }\Eprint {http://arxiv.org/abs/2103.06674}
  {arXiv:2103.06674} \BibitemShut {NoStop}%
\bibitem [{\citenamefont {Nomura}\ and\ \citenamefont
  {Arita}()}]{arita21review}%
  \BibitemOpen
  \bibfield  {author} {\bibinfo {author} {\bibfnamefont {Y.}~\bibnamefont
  {Nomura}}\ and\ \bibinfo {author} {\bibfnamefont {R.}~\bibnamefont {Arita}},\
  }\href@noop {} {\ }\Eprint {http://arxiv.org/abs/2107.12923}
  {arXiv:2107.12923} \BibitemShut {NoStop}%
\bibitem [{\citenamefont {{Li}}\ \emph {et~al.}(2019)\citenamefont {{Li}},
  \citenamefont {{Lee}}, \citenamefont {{Wang}}, \citenamefont {{Osada}},
  \citenamefont {{Crossley}}, \citenamefont {{Lee}}, \citenamefont {{Cui}},
  \citenamefont {{Hikita}},\ and\ \citenamefont {{Hwang}}}]{hwang19a}%
  \BibitemOpen
  \bibfield  {author} {\bibinfo {author} {\bibfnamefont {D.}~\bibnamefont
  {{Li}}}, \bibinfo {author} {\bibfnamefont {K.}~\bibnamefont {{Lee}}},
  \bibinfo {author} {\bibfnamefont {B.~Y.}\ \bibnamefont {{Wang}}}, \bibinfo
  {author} {\bibfnamefont {M.}~\bibnamefont {{Osada}}}, \bibinfo {author}
  {\bibfnamefont {S.}~\bibnamefont {{Crossley}}}, \bibinfo {author}
  {\bibfnamefont {H.~R.}\ \bibnamefont {{Lee}}}, \bibinfo {author}
  {\bibfnamefont {Y.}~\bibnamefont {{Cui}}}, \bibinfo {author} {\bibfnamefont
  {Y.}~\bibnamefont {{Hikita}}}, \ and\ \bibinfo {author} {\bibfnamefont
  {H.~Y.}\ \bibnamefont {{Hwang}}},\ }\href {\doibase
  10.1038/s41586-019-1496-5} {\bibfield  {journal} {\bibinfo  {journal} {\nat}\
  }\textbf {\bibinfo {volume} {572}},\ \bibinfo {pages} {624} (\bibinfo {year}
  {2019})}\BibitemShut {NoStop}%
\bibitem [{\citenamefont {Choi}\ \emph {et~al.}(2020)\citenamefont {Choi},
  \citenamefont {Lee},\ and\ \citenamefont {Pickett}}]{choi20_1}%
  \BibitemOpen
  \bibfield  {author} {\bibinfo {author} {\bibfnamefont {M.-Y.}\ \bibnamefont
  {Choi}}, \bibinfo {author} {\bibfnamefont {K.-W.}\ \bibnamefont {Lee}}, \
  and\ \bibinfo {author} {\bibfnamefont {W.~E.}\ \bibnamefont {Pickett}},\
  }\href {\doibase 10.1103/PhysRevB.101.020503} {\bibfield  {journal} {\bibinfo
   {journal} {Phys. Rev. B}\ }\textbf {\bibinfo {volume} {101}},\ \bibinfo
  {pages} {020503} (\bibinfo {year} {2020})}\BibitemShut {NoStop}%
\bibitem [{\citenamefont {Si}\ \emph {et~al.}(2020)\citenamefont {Si},
  \citenamefont {Xiao}, \citenamefont {Kaufmann}, \citenamefont {Tomczak},
  \citenamefont {Lu}, \citenamefont {Zhong},\ and\ \citenamefont
  {Held}}]{held20topotaticH}%
  \BibitemOpen
  \bibfield  {author} {\bibinfo {author} {\bibfnamefont {L.}~\bibnamefont
  {Si}}, \bibinfo {author} {\bibfnamefont {W.}~\bibnamefont {Xiao}}, \bibinfo
  {author} {\bibfnamefont {J.}~\bibnamefont {Kaufmann}}, \bibinfo {author}
  {\bibfnamefont {J.~M.}\ \bibnamefont {Tomczak}}, \bibinfo {author}
  {\bibfnamefont {Y.}~\bibnamefont {Lu}}, \bibinfo {author} {\bibfnamefont
  {Z.}~\bibnamefont {Zhong}}, \ and\ \bibinfo {author} {\bibfnamefont
  {K.}~\bibnamefont {Held}},\ }\href {\doibase 10.1103/PhysRevLett.124.166402}
  {\bibfield  {journal} {\bibinfo  {journal} {Phys. Rev. Lett.}\ }\textbf
  {\bibinfo {volume} {124}},\ \bibinfo {pages} {166402} (\bibinfo {year}
  {2020})}\BibitemShut {NoStop}%
\bibitem [{\citenamefont {{Bernardini}}\ and\ \citenamefont
  {{Cano}}(2020)}]{cano20c}%
  \BibitemOpen
  \bibfield  {author} {\bibinfo {author} {\bibfnamefont {F.}~\bibnamefont
  {{Bernardini}}}\ and\ \bibinfo {author} {\bibfnamefont {A.}~\bibnamefont
  {{Cano}}},\ }\href {\doibase 10.1088/2515-7639/ab9d0f} {\bibfield  {journal}
  {\bibinfo  {journal} {J. Phys. Mater.}\ }\textbf {\bibinfo {volume} {3}},\
  \bibinfo {eid} {03LT01} (\bibinfo {year} {2020})}\BibitemShut {NoStop}%
\bibitem [{\citenamefont {Osada}\ \emph {et~al.}(2021)\citenamefont {Osada},
  \citenamefont {Wang}, \citenamefont {Goodge}, \citenamefont {Harvey},
  \citenamefont {Lee}, \citenamefont {Li}, \citenamefont {Kourkoutis},\ and\
  \citenamefont {Hwang}}]{hwang21-La}%
  \BibitemOpen
  \bibfield  {author} {\bibinfo {author} {\bibfnamefont {M.}~\bibnamefont
  {Osada}}, \bibinfo {author} {\bibfnamefont {B.~Y.}\ \bibnamefont {Wang}},
  \bibinfo {author} {\bibfnamefont {B.~H.}\ \bibnamefont {Goodge}}, \bibinfo
  {author} {\bibfnamefont {S.~P.}\ \bibnamefont {Harvey}}, \bibinfo {author}
  {\bibfnamefont {K.}~\bibnamefont {Lee}}, \bibinfo {author} {\bibfnamefont
  {D.}~\bibnamefont {Li}}, \bibinfo {author} {\bibfnamefont {L.~F.}\
  \bibnamefont {Kourkoutis}}, \ and\ \bibinfo {author} {\bibfnamefont {H.~Y.}\
  \bibnamefont {Hwang}},\ }\href {\doibase 10.1002/adma.202104083} {\bibfield
  {journal} {\bibinfo  {journal} {Advanced Materials}\ ,\ \bibinfo {pages}
  {2104083}} (\bibinfo {year} {2021})}\BibitemShut {NoStop}%
\bibitem [{\citenamefont {{Zeng}}\ \emph {et~al.}()\citenamefont {{Zeng}},
  \citenamefont {{Li}}, \citenamefont {{Chow}}, \citenamefont {{Cao}},
  \citenamefont {{Zhang}}, \citenamefont {{Tang}}, \citenamefont {{Yin}},
  \citenamefont {{Lim}}, \citenamefont {{Hu}}, \citenamefont {{Yang}},\ and\
  \citenamefont {{Ariando}}}]{ariando21-La}%
  \BibitemOpen
  \bibfield  {author} {\bibinfo {author} {\bibfnamefont {S.~W.}\ \bibnamefont
  {{Zeng}}}, \bibinfo {author} {\bibfnamefont {C.~J.}\ \bibnamefont {{Li}}},
  \bibinfo {author} {\bibfnamefont {L.~E.}\ \bibnamefont {{Chow}}}, \bibinfo
  {author} {\bibfnamefont {Y.}~\bibnamefont {{Cao}}}, \bibinfo {author}
  {\bibfnamefont {Z.~T.}\ \bibnamefont {{Zhang}}}, \bibinfo {author}
  {\bibfnamefont {C.~S.}\ \bibnamefont {{Tang}}}, \bibinfo {author}
  {\bibfnamefont {X.~M.}\ \bibnamefont {{Yin}}}, \bibinfo {author}
  {\bibfnamefont {Z.~S.}\ \bibnamefont {{Lim}}}, \bibinfo {author}
  {\bibfnamefont {J.~X.}\ \bibnamefont {{Hu}}}, \bibinfo {author}
  {\bibfnamefont {P.}~\bibnamefont {{Yang}}}, \ and\ \bibinfo {author}
  {\bibfnamefont {A.}~\bibnamefont {{Ariando}}},\ }\href@noop {} {\ }\Eprint
  {http://arxiv.org/abs/2105.13492} {arXiv:2105.13492} \BibitemShut {NoStop}%
\bibitem [{\citenamefont {Osada}\ \emph {et~al.}(2020)\citenamefont {Osada},
  \citenamefont {Wang}, \citenamefont {Goodge}, \citenamefont {Lee},
  \citenamefont {Yoon}, \citenamefont {Sakuma}, \citenamefont {Li},
  \citenamefont {Miura}, \citenamefont {Kourkoutis},\ and\ \citenamefont
  {Hwang}}]{hwang20Pr-a}%
  \BibitemOpen
  \bibfield  {author} {\bibinfo {author} {\bibfnamefont {M.}~\bibnamefont
  {Osada}}, \bibinfo {author} {\bibfnamefont {B.~Y.}\ \bibnamefont {Wang}},
  \bibinfo {author} {\bibfnamefont {B.~H.}\ \bibnamefont {Goodge}}, \bibinfo
  {author} {\bibfnamefont {K.}~\bibnamefont {Lee}}, \bibinfo {author}
  {\bibfnamefont {H.}~\bibnamefont {Yoon}}, \bibinfo {author} {\bibfnamefont
  {K.}~\bibnamefont {Sakuma}}, \bibinfo {author} {\bibfnamefont
  {D.}~\bibnamefont {Li}}, \bibinfo {author} {\bibfnamefont {M.}~\bibnamefont
  {Miura}}, \bibinfo {author} {\bibfnamefont {L.~F.}\ \bibnamefont
  {Kourkoutis}}, \ and\ \bibinfo {author} {\bibfnamefont {H.~Y.}\ \bibnamefont
  {Hwang}},\ }\href {http://dx.doi.org/10.1021/acs.nanolett.0c01392} {\bibfield
   {journal} {\bibinfo  {journal} {Nano Letters}\ }\textbf {\bibinfo {volume}
  {20}},\ \bibinfo {pages} {5735–5740} (\bibinfo {year} {2020})}\BibitemShut
  {NoStop}%
\bibitem [{\citenamefont {Lin}\ \emph {et~al.}(2022)\citenamefont {Lin},
  \citenamefont {Gawryluk}, \citenamefont {Klein}, \citenamefont {Huangfu},
  \citenamefont {Pomjakushina}, \citenamefont {von Rohr},\ and\ \citenamefont
  {Schilling}}]{schilling22-njphy}%
  \BibitemOpen
  \bibfield  {author} {\bibinfo {author} {\bibfnamefont {H.}~\bibnamefont
  {Lin}}, \bibinfo {author} {\bibfnamefont {D.~J.}\ \bibnamefont {Gawryluk}},
  \bibinfo {author} {\bibfnamefont {Y.~M.}\ \bibnamefont {Klein}}, \bibinfo
  {author} {\bibfnamefont {S.}~\bibnamefont {Huangfu}}, \bibinfo {author}
  {\bibfnamefont {E.}~\bibnamefont {Pomjakushina}}, \bibinfo {author}
  {\bibfnamefont {F.}~\bibnamefont {von Rohr}}, \ and\ \bibinfo {author}
  {\bibfnamefont {A.}~\bibnamefont {Schilling}},\ }\href {\doibase
  10.1088/1367-2630/ac465e} {\bibfield  {journal} {\bibinfo  {journal} {New
  Journal of Physics}\ }\textbf {\bibinfo {volume} {24}},\ \bibinfo {pages}
  {013022} (\bibinfo {year} {2022})}\BibitemShut {NoStop}%
\bibitem [{\citenamefont {{He}}\ \emph {et~al.}()\citenamefont {{He}},
  \citenamefont {{Ming}}, \citenamefont {{Li}}, \citenamefont {{Zhu}},
  \citenamefont {{Si}},\ and\ \citenamefont {{Wen}}}]{sm-112}%
  \BibitemOpen
  \bibfield  {author} {\bibinfo {author} {\bibfnamefont {C.}~\bibnamefont
  {{He}}}, \bibinfo {author} {\bibfnamefont {X.}~\bibnamefont {{Ming}}},
  \bibinfo {author} {\bibfnamefont {Q.}~\bibnamefont {{Li}}}, \bibinfo {author}
  {\bibfnamefont {X.}~\bibnamefont {{Zhu}}}, \bibinfo {author} {\bibfnamefont
  {J.}~\bibnamefont {{Si}}}, \ and\ \bibinfo {author} {\bibfnamefont {H.-H.}\
  \bibnamefont {{Wen}}},\ }\href@noop {} {\ }\Eprint
  {http://arxiv.org/abs/2010.11777} {arXiv:2010.11777} \BibitemShut {NoStop}%
\bibitem [{\citenamefont {Botana}\ \emph {et~al.}(2017)\citenamefont {Botana},
  \citenamefont {Pardo},\ and\ \citenamefont {Norman}}]{botanaprm}%
  \BibitemOpen
  \bibfield  {author} {\bibinfo {author} {\bibfnamefont {A.~S.}\ \bibnamefont
  {Botana}}, \bibinfo {author} {\bibfnamefont {V.}~\bibnamefont {Pardo}}, \
  and\ \bibinfo {author} {\bibfnamefont {M.~R.}\ \bibnamefont {Norman}},\
  }\href {\doibase 10.1103/PhysRevMaterials.1.021801} {\bibfield  {journal}
  {\bibinfo  {journal} {Phys. Rev. Materials}\ }\textbf {\bibinfo {volume}
  {1}},\ \bibinfo {pages} {021801} (\bibinfo {year} {2017})}\BibitemShut
  {NoStop}%
\bibitem [{\citenamefont {{Kapeghian}}\ and\ \citenamefont
  {{Botana}}(2020)}]{botana20magnetism}%
  \BibitemOpen
  \bibfield  {author} {\bibinfo {author} {\bibfnamefont {J.}~\bibnamefont
  {{Kapeghian}}}\ and\ \bibinfo {author} {\bibfnamefont {A.~S.}\ \bibnamefont
  {{Botana}}},\ }\href {https://link.aps.org/doi/10.1103/PhysRevB.102.205130}
  {\bibfield  {journal} {\bibinfo  {journal} {Phys. Rev. B}\ }\textbf {\bibinfo
  {volume} {102}},\ \bibinfo {pages} {205130} (\bibinfo {year}
  {2020})}\BibitemShut {NoStop}%
\bibitem [{\citenamefont {Been}\ \emph {et~al.}(2021)\citenamefont {Been},
  \citenamefont {Lee}, \citenamefont {Hwang}, \citenamefont {Cui},
  \citenamefont {Zaanen}, \citenamefont {Devereaux}, \citenamefont {Moritz},\
  and\ \citenamefont {Jia}}]{jia-prx21}%
  \BibitemOpen
  \bibfield  {author} {\bibinfo {author} {\bibfnamefont {E.}~\bibnamefont
  {Been}}, \bibinfo {author} {\bibfnamefont {W.-S.}\ \bibnamefont {Lee}},
  \bibinfo {author} {\bibfnamefont {H.~Y.}\ \bibnamefont {Hwang}}, \bibinfo
  {author} {\bibfnamefont {Y.}~\bibnamefont {Cui}}, \bibinfo {author}
  {\bibfnamefont {J.}~\bibnamefont {Zaanen}}, \bibinfo {author} {\bibfnamefont
  {T.}~\bibnamefont {Devereaux}}, \bibinfo {author} {\bibfnamefont
  {B.}~\bibnamefont {Moritz}}, \ and\ \bibinfo {author} {\bibfnamefont
  {C.}~\bibnamefont {Jia}},\ }\href {\doibase 10.1103/PhysRevX.11.011050}
  {\bibfield  {journal} {\bibinfo  {journal} {Phys. Rev. X}\ }\textbf {\bibinfo
  {volume} {11}},\ \bibinfo {pages} {011050} (\bibinfo {year}
  {2021})}\BibitemShut {NoStop}%
\bibitem [{\citenamefont {Hirayama}\ \emph {et~al.}(2020)\citenamefont
  {Hirayama}, \citenamefont {Tadano}, \citenamefont {Nomura},\ and\
  \citenamefont {Arita}}]{arita20prb}%
  \BibitemOpen
  \bibfield  {author} {\bibinfo {author} {\bibfnamefont {M.}~\bibnamefont
  {Hirayama}}, \bibinfo {author} {\bibfnamefont {T.}~\bibnamefont {Tadano}},
  \bibinfo {author} {\bibfnamefont {Y.}~\bibnamefont {Nomura}}, \ and\ \bibinfo
  {author} {\bibfnamefont {R.}~\bibnamefont {Arita}},\ }\href {\doibase
  10.1103/PhysRevB.101.075107} {\bibfield  {journal} {\bibinfo  {journal}
  {Phys. Rev. B}\ }\textbf {\bibinfo {volume} {101}},\ \bibinfo {pages}
  {075107} (\bibinfo {year} {2020})}\BibitemShut {NoStop}%
\bibitem [{\citenamefont {Bernardini}\ \emph {et~al.}(2020)\citenamefont
  {Bernardini}, \citenamefont {Olevano}, \citenamefont {Blase},\ and\
  \citenamefont {Cano}}]{cano20d}%
  \BibitemOpen
  \bibfield  {author} {\bibinfo {author} {\bibfnamefont {F.}~\bibnamefont
  {Bernardini}}, \bibinfo {author} {\bibfnamefont {V.}~\bibnamefont {Olevano}},
  \bibinfo {author} {\bibfnamefont {X.}~\bibnamefont {Blase}}, \ and\ \bibinfo
  {author} {\bibfnamefont {A.}~\bibnamefont {Cano}},\ }\href {\doibase
  10.1088/2515-7639/ab885d} {\bibfield  {journal} {\bibinfo  {journal} {J.
  Phys. Mater.}\ }\textbf {\bibinfo {volume} {3}},\ \bibinfo {pages} {035003}
  (\bibinfo {year} {2020})}\BibitemShut {NoStop}%
\bibitem [{\citenamefont {Bernardini}\ \emph {et~al.}(2021)\citenamefont
  {Bernardini}, \citenamefont {Demourgues},\ and\ \citenamefont
  {Cano}}]{cano21-prm}%
  \BibitemOpen
  \bibfield  {author} {\bibinfo {author} {\bibfnamefont {F.}~\bibnamefont
  {Bernardini}}, \bibinfo {author} {\bibfnamefont {A.}~\bibnamefont
  {Demourgues}}, \ and\ \bibinfo {author} {\bibfnamefont {A.}~\bibnamefont
  {Cano}},\ }\href {\doibase 10.1103/PhysRevMaterials.5.L061801} {\bibfield
  {journal} {\bibinfo  {journal} {Phys. Rev. Materials}\ }\textbf {\bibinfo
  {volume} {5}},\ \bibinfo {pages} {L061801} (\bibinfo {year}
  {2021})}\BibitemShut {NoStop}%
\bibitem [{\citenamefont {Li}\ \emph {et~al.}(2020)\citenamefont {Li},
  \citenamefont {Guo}, \citenamefont {Zhang}, \citenamefont {Song},
  \citenamefont {Gao}, \citenamefont {Gu},\ and\ \citenamefont
  {Nie}}]{higher_order}%
  \BibitemOpen
  \bibfield  {author} {\bibinfo {author} {\bibfnamefont {Z.}~\bibnamefont
  {Li}}, \bibinfo {author} {\bibfnamefont {W.}~\bibnamefont {Guo}}, \bibinfo
  {author} {\bibfnamefont {T.~T.}\ \bibnamefont {Zhang}}, \bibinfo {author}
  {\bibfnamefont {J.~H.}\ \bibnamefont {Song}}, \bibinfo {author}
  {\bibfnamefont {T.~Y.}\ \bibnamefont {Gao}}, \bibinfo {author} {\bibfnamefont
  {Z.~B.}\ \bibnamefont {Gu}}, \ and\ \bibinfo {author} {\bibfnamefont {Y.~F.}\
  \bibnamefont {Nie}},\ }\href {\doibase 10.1063/5.0018934} {\bibfield
  {journal} {\bibinfo  {journal} {APL Materials}\ }\textbf {\bibinfo {volume}
  {8}},\ \bibinfo {pages} {091112} (\bibinfo {year} {2020})}\BibitemShut
  {NoStop}%
\bibitem [{\citenamefont {{Pan}}\ \emph {et~al.}()\citenamefont {{Pan}},
  \citenamefont {{Ferenc Segedin}}, \citenamefont {{LaBollita}}, \citenamefont
  {{Song}}, \citenamefont {{Nica}}, \citenamefont {{Goodge}}, \citenamefont
  {{Pierce}}, \citenamefont {{Doyle}}, \citenamefont {{Novakov}}, \citenamefont
  {{C{\'o}rdova Carrizales}}, \citenamefont {{N'Diaye}}, \citenamefont
  {{Shafer}}, \citenamefont {{Paik}}, \citenamefont {{Heron}}, \citenamefont
  {{Mason}}, \citenamefont {{Yacoby}}, \citenamefont {{Kourkoutis}},
  \citenamefont {{Erten}}, \citenamefont {{Brooks}}, \citenamefont {{Botana}},\
  and\ \citenamefont {{Mundy}}}]{mundy21}%
  \BibitemOpen
  \bibfield  {author} {\bibinfo {author} {\bibfnamefont {G.~A.}\ \bibnamefont
  {{Pan}}}, \bibinfo {author} {\bibfnamefont {D.}~\bibnamefont {{Ferenc
  Segedin}}}, \bibinfo {author} {\bibfnamefont {H.}~\bibnamefont
  {{LaBollita}}}, \bibinfo {author} {\bibfnamefont {Q.}~\bibnamefont {{Song}}},
  \bibinfo {author} {\bibfnamefont {E.~M.}\ \bibnamefont {{Nica}}}, \bibinfo
  {author} {\bibfnamefont {B.~H.}\ \bibnamefont {{Goodge}}}, \bibinfo {author}
  {\bibfnamefont {A.~T.}\ \bibnamefont {{Pierce}}}, \bibinfo {author}
  {\bibfnamefont {S.}~\bibnamefont {{Doyle}}}, \bibinfo {author} {\bibfnamefont
  {S.}~\bibnamefont {{Novakov}}}, \bibinfo {author} {\bibfnamefont
  {D.}~\bibnamefont {{C{\'o}rdova Carrizales}}}, \bibinfo {author}
  {\bibfnamefont {A.~T.}\ \bibnamefont {{N'Diaye}}}, \bibinfo {author}
  {\bibfnamefont {P.}~\bibnamefont {{Shafer}}}, \bibinfo {author}
  {\bibfnamefont {H.}~\bibnamefont {{Paik}}}, \bibinfo {author} {\bibfnamefont
  {J.~T.}\ \bibnamefont {{Heron}}}, \bibinfo {author} {\bibfnamefont {J.~A.}\
  \bibnamefont {{Mason}}}, \bibinfo {author} {\bibfnamefont {A.}~\bibnamefont
  {{Yacoby}}}, \bibinfo {author} {\bibfnamefont {L.~F.}\ \bibnamefont
  {{Kourkoutis}}}, \bibinfo {author} {\bibfnamefont {O.}~\bibnamefont
  {{Erten}}}, \bibinfo {author} {\bibfnamefont {C.~M.}\ \bibnamefont
  {{Brooks}}}, \bibinfo {author} {\bibfnamefont {A.~S.}\ \bibnamefont
  {{Botana}}}, \ and\ \bibinfo {author} {\bibfnamefont {J.~A.}\ \bibnamefont
  {{Mundy}}},\ }\href@noop {} {\ }\Eprint {http://arxiv.org/abs/2109.09726}
  {arXiv:2109.09726} \BibitemShut {NoStop}%
\bibitem [{\citenamefont {{Mercy}}\ \emph {et~al.}(2017)\citenamefont
  {{Mercy}}, \citenamefont {{Bieder}}, \citenamefont {{Ì{\~ n}iguez}},\ and\
  \citenamefont {{Ghosez}}}]{ghosez17}%
  \BibitemOpen
  \bibfield  {author} {\bibinfo {author} {\bibfnamefont {A.}~\bibnamefont
  {{Mercy}}}, \bibinfo {author} {\bibfnamefont {J.}~\bibnamefont {{Bieder}}},
  \bibinfo {author} {\bibfnamefont {J.}~\bibnamefont {{Ì{\~ n}iguez}}}, \ and\
  \bibinfo {author} {\bibfnamefont {P.}~\bibnamefont {{Ghosez}}},\ }\href
  {\doibase 10.1038/s41467-017-01811-x} {\bibfield  {journal} {\bibinfo
  {journal} {Nature Comm.}\ }\textbf {\bibinfo {volume} {8}},\ \bibinfo {eid}
  {1677} (\bibinfo {year} {2017})}\BibitemShut {NoStop}%
\bibitem [{\citenamefont {{Varignon}}\ \emph {et~al.}(2017)\citenamefont
  {{Varignon}}, \citenamefont {{Grisolia}}, \citenamefont {{{\'I}{\~n}iguez}},
  \citenamefont {{Barth{\'e}l{\'e}my}},\ and\ \citenamefont
  {{Bibes}}}]{bibes17}%
  \BibitemOpen
  \bibfield  {author} {\bibinfo {author} {\bibfnamefont {J.}~\bibnamefont
  {{Varignon}}}, \bibinfo {author} {\bibfnamefont {M.~N.}\ \bibnamefont
  {{Grisolia}}}, \bibinfo {author} {\bibfnamefont {J.}~\bibnamefont
  {{{\'I}{\~n}iguez}}}, \bibinfo {author} {\bibfnamefont {A.}~\bibnamefont
  {{Barth{\'e}l{\'e}my}}}, \ and\ \bibinfo {author} {\bibfnamefont
  {M.}~\bibnamefont {{Bibes}}},\ }\href {\doibase 10.1038/s41535-017-0024-9}
  {\bibfield  {journal} {\bibinfo  {journal} {npj Quantum Materials}\ }\textbf
  {\bibinfo {volume} {2}},\ \bibinfo {eid} {21} (\bibinfo {year}
  {2017})}\BibitemShut {NoStop}%
\bibitem [{\citenamefont {Gawryluk}\ \emph {et~al.}(2019)\citenamefont
  {Gawryluk}, \citenamefont {Klein}, \citenamefont {Shang}, \citenamefont
  {Sheptyakov}, \citenamefont {Keller}, \citenamefont {Casati}, \citenamefont
  {Lacorre}, \citenamefont {Fern\'andez-D\'{\i}az}, \citenamefont
  {Rodr\'{\i}guez-Carvajal},\ and\ \citenamefont {Medarde}}]{medarde19}%
  \BibitemOpen
  \bibfield  {author} {\bibinfo {author} {\bibfnamefont {D.~J.}\ \bibnamefont
  {Gawryluk}}, \bibinfo {author} {\bibfnamefont {Y.~M.}\ \bibnamefont {Klein}},
  \bibinfo {author} {\bibfnamefont {T.}~\bibnamefont {Shang}}, \bibinfo
  {author} {\bibfnamefont {D.}~\bibnamefont {Sheptyakov}}, \bibinfo {author}
  {\bibfnamefont {L.}~\bibnamefont {Keller}}, \bibinfo {author} {\bibfnamefont
  {N.}~\bibnamefont {Casati}}, \bibinfo {author} {\bibfnamefont
  {P.}~\bibnamefont {Lacorre}}, \bibinfo {author} {\bibfnamefont {M.~T.}\
  \bibnamefont {Fern\'andez-D\'{\i}az}}, \bibinfo {author} {\bibfnamefont
  {J.}~\bibnamefont {Rodr\'{\i}guez-Carvajal}}, \ and\ \bibinfo {author}
  {\bibfnamefont {M.}~\bibnamefont {Medarde}},\ }\href {\doibase
  10.1103/PhysRevB.100.205137} {\bibfield  {journal} {\bibinfo  {journal}
  {Phys. Rev. B}\ }\textbf {\bibinfo {volume} {100}},\ \bibinfo {pages}
  {205137} (\bibinfo {year} {2019})}\BibitemShut {NoStop}%
\bibitem [{\citenamefont {Giannozzi}\ and\ \citenamefont {{\it et
  al.}}(2009)}]{QE}%
  \BibitemOpen
  \bibfield  {author} {\bibinfo {author} {\bibfnamefont {P.}~\bibnamefont
  {Giannozzi}}\ and\ \bibinfo {author} {\bibnamefont {{\it et al.}}},\ }\href
  {\doibase 10.1088/0953-8984/21/39/395502} {\bibfield  {journal} {\bibinfo
  {journal} {J. Phys.: Condens. Matter}\ }\textbf {\bibinfo {volume} {21}},\
  \bibinfo {pages} {395502} (\bibinfo {year} {2009})}\BibitemShut {NoStop}%
\bibitem [{\citenamefont {Hamann}(2013)}]{ONCVPSP}%
  \BibitemOpen
  \bibfield  {author} {\bibinfo {author} {\bibfnamefont {D.~R.}\ \bibnamefont
  {Hamann}},\ }\href {\doibase 10.1103/PhysRevB.88.085117} {\bibfield
  {journal} {\bibinfo  {journal} {Phys. Rev. B}\ }\textbf {\bibinfo {volume}
  {88}},\ \bibinfo {pages} {085117} (\bibinfo {year} {2013})}\BibitemShut
  {NoStop}%
\bibitem [{\citenamefont {{van Setten}}\ \emph {et~al.}(2018)\citenamefont
  {{van Setten}}, \citenamefont {Giantomassi}, \citenamefont {Bousquet},
  \citenamefont {Verstraete}, \citenamefont {Hamann}, \citenamefont {Gonze},\
  and\ \citenamefont {Rignanese}}]{dojo}%
  \BibitemOpen
  \bibfield  {author} {\bibinfo {author} {\bibfnamefont {M.}~\bibnamefont {{van
  Setten}}}, \bibinfo {author} {\bibfnamefont {M.}~\bibnamefont {Giantomassi}},
  \bibinfo {author} {\bibfnamefont {E.}~\bibnamefont {Bousquet}}, \bibinfo
  {author} {\bibfnamefont {M.}~\bibnamefont {Verstraete}}, \bibinfo {author}
  {\bibfnamefont {D.}~\bibnamefont {Hamann}}, \bibinfo {author} {\bibfnamefont
  {X.}~\bibnamefont {Gonze}}, \ and\ \bibinfo {author} {\bibfnamefont {G.-M.}\
  \bibnamefont {Rignanese}},\ }\href {\doibase
  https://doi.org/10.1016/j.cpc.2018.01.012} {\bibfield  {journal} {\bibinfo
  {journal} {Comput. Phys. Commun.}\ }\textbf {\bibinfo {volume} {226}},\
  \bibinfo {pages} {39} (\bibinfo {year} {2018})}\BibitemShut {NoStop}%
\bibitem [{\citenamefont {Perdew}\ \emph {et~al.}(2008)\citenamefont {Perdew},
  \citenamefont {Ruzsinszky}, \citenamefont {Csonka}, \citenamefont {Vydrov},
  \citenamefont {Scuseria}, \citenamefont {Constantin}, \citenamefont {Zhou},\
  and\ \citenamefont {Burke}}]{PBEsol}%
  \BibitemOpen
  \bibfield  {author} {\bibinfo {author} {\bibfnamefont {J.~P.}\ \bibnamefont
  {Perdew}}, \bibinfo {author} {\bibfnamefont {A.}~\bibnamefont {Ruzsinszky}},
  \bibinfo {author} {\bibfnamefont {G.~I.}\ \bibnamefont {Csonka}}, \bibinfo
  {author} {\bibfnamefont {O.~A.}\ \bibnamefont {Vydrov}}, \bibinfo {author}
  {\bibfnamefont {G.~E.}\ \bibnamefont {Scuseria}}, \bibinfo {author}
  {\bibfnamefont {L.~A.}\ \bibnamefont {Constantin}}, \bibinfo {author}
  {\bibfnamefont {X.}~\bibnamefont {Zhou}}, \ and\ \bibinfo {author}
  {\bibfnamefont {K.}~\bibnamefont {Burke}},\ }\href {\doibase
  10.1103/PhysRevLett.100.136406} {\bibfield  {journal} {\bibinfo  {journal}
  {Phys. Rev. Lett.}\ }\textbf {\bibinfo {volume} {100}},\ \bibinfo {pages}
  {136406} (\bibinfo {year} {2008})}\BibitemShut {NoStop}%
\bibitem [{\citenamefont {Prandini}\ \emph {et~al.}(2018)\citenamefont
  {Prandini}, \citenamefont {Marrazzo}, \citenamefont {Castelli}, \citenamefont
  {Mounet},\ and\ \citenamefont {Marzari}}]{SSSP}%
  \BibitemOpen
  \bibfield  {author} {\bibinfo {author} {\bibfnamefont {G.}~\bibnamefont
  {Prandini}}, \bibinfo {author} {\bibfnamefont {A.}~\bibnamefont {Marrazzo}},
  \bibinfo {author} {\bibfnamefont {I.~E.}\ \bibnamefont {Castelli}}, \bibinfo
  {author} {\bibfnamefont {N.}~\bibnamefont {Mounet}}, \ and\ \bibinfo {author}
  {\bibfnamefont {N.}~\bibnamefont {Marzari}},\ }\href {\doibase
  10.1038/s41524-018-0127-2} {\bibfield  {journal} {\bibinfo  {journal} {npj
  Computational Materials}\ }\textbf {\bibinfo {volume} {4}},\ \bibinfo {pages}
  {72} (\bibinfo {year} {2018})}\BibitemShut {NoStop}%
\bibitem [{\citenamefont {Baroni}\ \emph {et~al.}(1987)\citenamefont {Baroni},
  \citenamefont {Giannozzi},\ and\ \citenamefont {Testa}}]{DFPT}%
  \BibitemOpen
  \bibfield  {author} {\bibinfo {author} {\bibfnamefont {S.}~\bibnamefont
  {Baroni}}, \bibinfo {author} {\bibfnamefont {P.}~\bibnamefont {Giannozzi}}, \
  and\ \bibinfo {author} {\bibfnamefont {A.}~\bibnamefont {Testa}},\ }\href
  {\doibase 10.1103/PhysRevLett.58.1861} {\bibfield  {journal} {\bibinfo
  {journal} {Phys. Rev. Lett.}\ }\textbf {\bibinfo {volume} {58}},\ \bibinfo
  {pages} {1861} (\bibinfo {year} {1987})}\BibitemShut {NoStop}%
\bibitem [{\citenamefont {Blaha}\ \emph {et~al.}()\citenamefont {Blaha},
  \citenamefont {Schwarz}, \citenamefont {Madsen}, \citenamefont {Kvasnicka},
  \citenamefont {Luitz}, \citenamefont {Laskowski}, \citenamefont {Tran},\ and\
  \citenamefont {Marks}}]{WIEN2k}%
  \BibitemOpen
  \bibfield  {author} {\bibinfo {author} {\bibfnamefont {P.}~\bibnamefont
  {Blaha}}, \bibinfo {author} {\bibfnamefont {K.}~\bibnamefont {Schwarz}},
  \bibinfo {author} {\bibfnamefont {G.}~\bibnamefont {Madsen}}, \bibinfo
  {author} {\bibfnamefont {D.}~\bibnamefont {Kvasnicka}}, \bibinfo {author}
  {\bibfnamefont {J.}~\bibnamefont {Luitz}}, \bibinfo {author} {\bibfnamefont
  {R.}~\bibnamefont {Laskowski}}, \bibinfo {author} {\bibfnamefont
  {F.}~\bibnamefont {Tran}}, \ and\ \bibinfo {author} {\bibfnamefont {L.~D.}\
  \bibnamefont {Marks}},\ }\href@noop {} {\bibinfo  {journal} {{W}{I}{E}{N}2k,
  An Augmented Plane Wave + Local Orbitals Program for Calculating Crystal
  Properties (Karlheinz Schwarz, Techn. Universität Wien, Austria), 2018. ISBN
  3-9501031-1-2}\ }\BibitemShut {NoStop}%
\bibitem [{\citenamefont {Perdew}\ and\ \citenamefont {Zunger}(1981)}]{LDA}%
  \BibitemOpen
\bibfield  {journal} {  }\bibfield  {author} {\bibinfo {author} {\bibfnamefont
  {J.~P.}\ \bibnamefont {Perdew}}\ and\ \bibinfo {author} {\bibfnamefont
  {A.}~\bibnamefont {Zunger}},\ }\href {\doibase 10.1103/PhysRevB.23.5048}
  {\bibfield  {journal} {\bibinfo  {journal} {Phys. Rev. B}\ }\textbf {\bibinfo
  {volume} {23}},\ \bibinfo {pages} {5048} (\bibinfo {year}
  {1981})}\BibitemShut {NoStop}%
\bibitem [{\citenamefont {Mazin}\ \emph {et~al.}(2008)\citenamefont {Mazin},
  \citenamefont {Johannes}, \citenamefont {Boeri}, \citenamefont {Koepernik},\
  and\ \citenamefont {Singh}}]{mazin08-prb}%
  \BibitemOpen
  \bibfield  {author} {\bibinfo {author} {\bibfnamefont {I.~I.}\ \bibnamefont
  {Mazin}}, \bibinfo {author} {\bibfnamefont {M.~D.}\ \bibnamefont {Johannes}},
  \bibinfo {author} {\bibfnamefont {L.}~\bibnamefont {Boeri}}, \bibinfo
  {author} {\bibfnamefont {K.}~\bibnamefont {Koepernik}}, \ and\ \bibinfo
  {author} {\bibfnamefont {D.~J.}\ \bibnamefont {Singh}},\ }\href {\doibase
  10.1103/PhysRevB.78.085104} {\bibfield  {journal} {\bibinfo  {journal} {Phys.
  Rev. B}\ }\textbf {\bibinfo {volume} {78}},\ \bibinfo {pages} {085104}
  (\bibinfo {year} {2008})}\BibitemShut {NoStop}%
\bibitem [{\citenamefont {Puphal}\ \emph {et~al.}(2021)\citenamefont {Puphal},
  \citenamefont {Wu}, \citenamefont {Fürsich}, \citenamefont {Lee},
  \citenamefont {Pakdaman}, \citenamefont {Bruin}, \citenamefont {Nuss},
  \citenamefont {Suyolcu}, \citenamefont {van Aken}, \citenamefont {Keimer},
  \citenamefont {Isobe},\ and\ \citenamefont {Hepting}}]{hepting21-sciadv}%
  \BibitemOpen
  \bibfield  {author} {\bibinfo {author} {\bibfnamefont {P.}~\bibnamefont
  {Puphal}}, \bibinfo {author} {\bibfnamefont {Y.-M.}\ \bibnamefont {Wu}},
  \bibinfo {author} {\bibfnamefont {K.}~\bibnamefont {Fürsich}}, \bibinfo
  {author} {\bibfnamefont {H.}~\bibnamefont {Lee}}, \bibinfo {author}
  {\bibfnamefont {M.}~\bibnamefont {Pakdaman}}, \bibinfo {author}
  {\bibfnamefont {J.~A.~N.}\ \bibnamefont {Bruin}}, \bibinfo {author}
  {\bibfnamefont {J.}~\bibnamefont {Nuss}}, \bibinfo {author} {\bibfnamefont
  {Y.~E.}\ \bibnamefont {Suyolcu}}, \bibinfo {author} {\bibfnamefont {P.~A.}\
  \bibnamefont {van Aken}}, \bibinfo {author} {\bibfnamefont {B.}~\bibnamefont
  {Keimer}}, \bibinfo {author} {\bibfnamefont {M.}~\bibnamefont {Isobe}}, \
  and\ \bibinfo {author} {\bibfnamefont {M.}~\bibnamefont {Hepting}},\ }\href
  {\doibase 10.1126/sciadv.abl8091} {\bibfield  {journal} {\bibinfo  {journal}
  {Science Advances}\ }\textbf {\bibinfo {volume} {7}},\ \bibinfo {pages}
  {eabl8091} (\bibinfo {year} {2021})}\BibitemShut {NoStop}%
\bibitem [{\citenamefont {Wang}\ \emph {et~al.}(2020)\citenamefont {Wang},
  \citenamefont {Zheng}, \citenamefont {Krivyakina}, \citenamefont {Chmaissem},
  \citenamefont {Lopes}, \citenamefont {Lynn}, \citenamefont {Gallington},
  \citenamefont {Ren}, \citenamefont {Rosenkranz}, \citenamefont {Mitchell},\
  and\ \citenamefont {Phelan}}]{mitchell20bulk}%
  \BibitemOpen
  \bibfield  {author} {\bibinfo {author} {\bibfnamefont {B.-X.}\ \bibnamefont
  {Wang}}, \bibinfo {author} {\bibfnamefont {H.}~\bibnamefont {Zheng}},
  \bibinfo {author} {\bibfnamefont {E.}~\bibnamefont {Krivyakina}}, \bibinfo
  {author} {\bibfnamefont {O.}~\bibnamefont {Chmaissem}}, \bibinfo {author}
  {\bibfnamefont {P.~P.}\ \bibnamefont {Lopes}}, \bibinfo {author}
  {\bibfnamefont {J.~W.}\ \bibnamefont {Lynn}}, \bibinfo {author}
  {\bibfnamefont {L.~C.}\ \bibnamefont {Gallington}}, \bibinfo {author}
  {\bibfnamefont {Y.}~\bibnamefont {Ren}}, \bibinfo {author} {\bibfnamefont
  {S.}~\bibnamefont {Rosenkranz}}, \bibinfo {author} {\bibfnamefont {J.~F.}\
  \bibnamefont {Mitchell}}, \ and\ \bibinfo {author} {\bibfnamefont
  {D.}~\bibnamefont {Phelan}},\ }\href {\doibase
  10.1103/PhysRevMaterials.4.084409} {\bibfield  {journal} {\bibinfo  {journal}
  {Phys. Rev. Materials}\ }\textbf {\bibinfo {volume} {4}},\ \bibinfo {pages}
  {084409} (\bibinfo {year} {2020})}\BibitemShut {NoStop}%
\bibitem [{\citenamefont {Qin}\ \emph {et~al.}(2005)\citenamefont {Qin},
  \citenamefont {Liu}, \citenamefont {Yu}, \citenamefont {Bao}, \citenamefont
  {Li}, \citenamefont {Yu}, \citenamefont {Liu},\ and\ \citenamefont
  {Jin}}]{jin05-bulkmodulus}%
  \BibitemOpen
  \bibfield  {author} {\bibinfo {author} {\bibfnamefont {X.}~\bibnamefont
  {Qin}}, \bibinfo {author} {\bibfnamefont {Q.}~\bibnamefont {Liu}}, \bibinfo
  {author} {\bibfnamefont {Y.}~\bibnamefont {Yu}}, \bibinfo {author}
  {\bibfnamefont {Z.}~\bibnamefont {Bao}}, \bibinfo {author} {\bibfnamefont
  {F.}~\bibnamefont {Li}}, \bibinfo {author} {\bibfnamefont {R.}~\bibnamefont
  {Yu}}, \bibinfo {author} {\bibfnamefont {J.}~\bibnamefont {Liu}}, \ and\
  \bibinfo {author} {\bibfnamefont {C.}~\bibnamefont {Jin}},\ }\href {\doibase
  10.1016/j.stam.2005.06.012} {\bibfield  {journal} {\bibinfo  {journal} {Sci.
  Technol. Adv. Mater.}\ }\textbf {\bibinfo {volume} {6}},\ \bibinfo {pages}
  {828} (\bibinfo {year} {2005})}\BibitemShut {NoStop}%
\bibitem [{\citenamefont {{Wang}}\ \emph {et~al.}()\citenamefont {{Wang}},
  \citenamefont {{Yang}}, \citenamefont {{Chen}}, \citenamefont {{Yang}},
  \citenamefont {{Zhang}}, \citenamefont {{Zhu}}, \citenamefont {{Uwatoko}},
  \citenamefont {{Dong}}, \citenamefont {{Jin}}, \citenamefont {{Sun}},\ and\
  \citenamefont {{Cheng}}}]{cheng21-pressure}%
  \BibitemOpen
  \bibfield  {author} {\bibinfo {author} {\bibfnamefont {N.~N.}\ \bibnamefont
  {{Wang}}}, \bibinfo {author} {\bibfnamefont {M.~W.}\ \bibnamefont {{Yang}}},
  \bibinfo {author} {\bibfnamefont {K.~Y.}\ \bibnamefont {{Chen}}}, \bibinfo
  {author} {\bibfnamefont {Z.}~\bibnamefont {{Yang}}}, \bibinfo {author}
  {\bibfnamefont {H.}~\bibnamefont {{Zhang}}}, \bibinfo {author} {\bibfnamefont
  {Z.~H.}\ \bibnamefont {{Zhu}}}, \bibinfo {author} {\bibfnamefont
  {Y.}~\bibnamefont {{Uwatoko}}}, \bibinfo {author} {\bibfnamefont {X.~L.}\
  \bibnamefont {{Dong}}}, \bibinfo {author} {\bibfnamefont {K.~J.}\
  \bibnamefont {{Jin}}}, \bibinfo {author} {\bibfnamefont {J.~P.}\ \bibnamefont
  {{Sun}}}, \ and\ \bibinfo {author} {\bibfnamefont {J.~G.}\ \bibnamefont
  {{Cheng}}},\ }\href@noop {} {\ }\Eprint {http://arxiv.org/abs/2109.12811}
  {arXiv:2109.12811} \BibitemShut {NoStop}%
\bibitem [{\citenamefont {Pauling}(1960)}]{pauling60}%
  \BibitemOpen
  \bibfield  {author} {\bibinfo {author} {\bibfnamefont {L.}~\bibnamefont
  {Pauling}},\ }\href@noop {} {\emph {\bibinfo {title} {The Nature of the
  Chemical Bond}}}\ (\bibinfo  {publisher} {Cornell University Press Ithaca,
  NY},\ \bibinfo {year} {1960})\BibitemShut {NoStop}%
\bibitem [{\citenamefont {Andersen}\ \emph {et~al.}(1996)\citenamefont
  {Andersen}, \citenamefont {Savrasov}, \citenamefont {Jepsen},\ and\
  \citenamefont {Liechtenstein}}]{liechtenstein96}%
  \BibitemOpen
  \bibfield  {author} {\bibinfo {author} {\bibfnamefont {O.~K.}\ \bibnamefont
  {Andersen}}, \bibinfo {author} {\bibfnamefont {S.~Y.}\ \bibnamefont
  {Savrasov}}, \bibinfo {author} {\bibfnamefont {O.}~\bibnamefont {Jepsen}}, \
  and\ \bibinfo {author} {\bibfnamefont {A.~I.}\ \bibnamefont
  {Liechtenstein}},\ }\href {\doibase 10.1007/bf00768402} {\bibfield  {journal}
  {\bibinfo  {journal} {J. Low Temp. Phys.}\ }\textbf {\bibinfo {volume}
  {105}},\ \bibinfo {pages} {285–304} (\bibinfo {year} {1996})}\BibitemShut
  {NoStop}%
\bibitem [{\citenamefont {Xia}\ \emph {et~al.}(2022)\citenamefont {Xia},
  \citenamefont {Wu}, \citenamefont {Chen},\ and\ \citenamefont
  {Chen}}]{chen21-dyn}%
  \BibitemOpen
  \bibfield  {author} {\bibinfo {author} {\bibfnamefont {C.}~\bibnamefont
  {Xia}}, \bibinfo {author} {\bibfnamefont {J.}~\bibnamefont {Wu}}, \bibinfo
  {author} {\bibfnamefont {Y.}~\bibnamefont {Chen}}, \ and\ \bibinfo {author}
  {\bibfnamefont {H.}~\bibnamefont {Chen}},\ }\href {\doibase
  10.1103/PhysRevB.105.115134} {\bibfield  {journal} {\bibinfo  {journal}
  {Phys. Rev. B}\ }\textbf {\bibinfo {volume} {105}},\ \bibinfo {pages}
  {115134} (\bibinfo {year} {2022})}\BibitemShut {NoStop}%
\bibitem [{\citenamefont {{Zhang}}\ \emph {et~al.}()\citenamefont {{Zhang}},
  \citenamefont {{Zhang}}, \citenamefont {{He}}, \citenamefont {{Wang}},\ and\
  \citenamefont {{Ghosez}}}]{ghosez21}%
  \BibitemOpen
  \bibfield  {author} {\bibinfo {author} {\bibfnamefont {Y.}~\bibnamefont
  {{Zhang}}}, \bibinfo {author} {\bibfnamefont {J.}~\bibnamefont {{Zhang}}},
  \bibinfo {author} {\bibfnamefont {X.}~\bibnamefont {{He}}}, \bibinfo {author}
  {\bibfnamefont {J.}~\bibnamefont {{Wang}}}, \ and\ \bibinfo {author}
  {\bibfnamefont {P.}~\bibnamefont {{Ghosez}}},\ }\href@noop {} {\ }\Eprint
  {http://arxiv.org/abs/2201.00709} {arXiv:2201.00709} \BibitemShut {NoStop}%
\bibitem [{\citenamefont {Carrasco~Álvarez}\ \emph {et~al.}()\citenamefont
  {Carrasco~Álvarez}, \citenamefont {Petit}, \citenamefont {Iglesias},
  \citenamefont {Prellier}, \citenamefont {Bibes},\ and\ \citenamefont
  {Varignon}}]{bibes21}%
  \BibitemOpen
  \bibfield  {author} {\bibinfo {author} {\bibfnamefont {A.~A.}\ \bibnamefont
  {Carrasco~Álvarez}}, \bibinfo {author} {\bibfnamefont {S.}~\bibnamefont
  {Petit}}, \bibinfo {author} {\bibfnamefont {L.}~\bibnamefont {Iglesias}},
  \bibinfo {author} {\bibfnamefont {W.}~\bibnamefont {Prellier}}, \bibinfo
  {author} {\bibfnamefont {M.}~\bibnamefont {Bibes}}, \ and\ \bibinfo {author}
  {\bibfnamefont {J.}~\bibnamefont {Varignon}},\ }\href@noop {} {\ }\Eprint
  {http://arxiv.org/abs/2112.02642} {arXiv:2112.02642} \BibitemShut {NoStop}%
\bibitem [{\citenamefont {Zhang}\ \emph {et~al.}(2021)\citenamefont {Zhang},
  \citenamefont {Lane}, \citenamefont {Singh}, \citenamefont {Nokelainen},
  \citenamefont {Barbiellini}, \citenamefont {Markiewicz}, \citenamefont
  {Bansil},\ and\ \citenamefont {Sun}}]{zhang21-commphys}%
  \BibitemOpen
  \bibfield  {author} {\bibinfo {author} {\bibfnamefont {R.}~\bibnamefont
  {Zhang}}, \bibinfo {author} {\bibfnamefont {C.}~\bibnamefont {Lane}},
  \bibinfo {author} {\bibfnamefont {B.}~\bibnamefont {Singh}}, \bibinfo
  {author} {\bibfnamefont {J.}~\bibnamefont {Nokelainen}}, \bibinfo {author}
  {\bibfnamefont {B.}~\bibnamefont {Barbiellini}}, \bibinfo {author}
  {\bibfnamefont {R.~S.}\ \bibnamefont {Markiewicz}}, \bibinfo {author}
  {\bibfnamefont {A.}~\bibnamefont {Bansil}}, \ and\ \bibinfo {author}
  {\bibfnamefont {J.}~\bibnamefont {Sun}},\ }\href@noop {} {\bibfield
  {journal} {\bibinfo  {journal} {Commun. Phys.}\ }\textbf {\bibinfo {volume}
  {4}},\ \bibinfo {pages} {1} (\bibinfo {year} {2021})}\BibitemShut {NoStop}%
\end{thebibliography}%

\end{document}